\documentclass[aps,pra,floatfix,twocolumn,superscriptaddress,reprint]{revtex4-1}
\usepackage{physics}
\usepackage{dcolumn}% Align table columns on decimal point
\usepackage{latexsym,epsfig,graphicx,dcolumn,subfigure,comment,ulem}
\usepackage{amsmath,eqnarray,amssymb,amsbsy}
\usepackage{xcolor,color}
\usepackage[colorlinks,urlcolor=blue,citecolor=blue]{hyperref}

\begin{document}

\title{Reverse strain-induced snake states in graphene nanoribbons}

\author{Cheng-yi Zuo}
\affiliation{Department of Physics, State Key Laboratory of 
	Precision Spectroscopy, School of Physics and Electronic 
	Science, East China Normal University, Shanghai 200241, China}
\author{Junjie Qi}
\affiliation{Beijing Academy of Quantum Information Sciences, 
	West Bld.3, No.10 Xibeiwang East Rd., Haidian District, 
	Beijing 100193, China}
\author{Tian-lun Lu}
\affiliation{Department of Physics, State Key Laboratory of 
	Precision Spectroscopy, School of Physics and Electronic 
	Science, East China Normal University, Shanghai 200241, China}
\author{Zhi-qiang Bao}
\email{zqbao@phy.ecnu.edu.cn}
\affiliation{Key Laboratory of Polar Materials and Devices (MOE), 
	Department of Electronics, East China Normal University,     
	Shanghai 200241, China}
\author{Yan Li}
\email{yli@phy.ecnu.edu.cn}
\affiliation{Department of Physics, State Key Laboratory of 
	Precision Spectroscopy, School of Physics and Electronic 
	Science, East China Normal University, Shanghai 200241, China}

\begin{abstract}

Strain can tailor the band structures and properties of graphene 
nanoribbons (GNRs) with the well-known emergent pseudo-magnetic fields 
and the corresponding pseudo-Landau levels (pLLs). We design 
one type of the zigzag GNR (ZGNR) with reverse strains, producing pseudo-magnetic 
fields with opposite signs in the lower and upper half planes. Therefore, 
electrons propagate along the interface as ``snake states", experiencing 
opposite Lorentz forces as they cross the zero field border line. By 
using the Landauer-B\"{u}ttiker formalism combined with the nonequilibrium 
Green's function method, the existence and robustness of the 
reverse strain-induced snake states are further studied. Furthermore, 
the realization of long-thought pure valley currents in monolayer graphene 
systems is also proposed in our device.
  
\end{abstract}

\maketitle

\section{Introduction}\label{sec:intro}

Graphene, a single-atom-layer carbon material, has the 
peculiar band structure with two nonequivalent Dirac points $K$ and 
$K'$, leading to a pseudo-spin degree of freedom, i.e., 
valley~\cite{grapheneRMP,Beenakker,Xiao1}. 
The field utilizing the valley degree of freedom is referred to as  
valleytronics~\cite{valleytronics}, which takes advantage of graphene 
and gapped 2 dimensional (2D) Dirac materials~\cite{Lensky,Xiao2,Chang,Slager}. Graphene 
nanoribbons (GNRs), cut from the graphene, are quasi-one-dimensional 
systems. There are two basic shapes for GNR edges — zigzag and 
armchair edges, and more specially, the zigzag GNRs (ZGNRs) 
support zero-energy flat bands of edge states~\cite{Dresselhaus,Wang,Bao,Lu,Rhim}. 
In our work, we mainly focus on the ZGNRs. Straintronics, 
a new discipline developed in recent decades, utilizes strain 
engineering methods and strain-induced physical effects to develop 
novel functional devices~\cite{straintronics1,straintronics2,straintronics3,Guo,Roy,Stuij}. 
Because graphene has the outstanding capability to sustain 
nondestructive reversible deformations up to high values $25\%$ to 
$27\%$~\cite{strength1,strength2,strength3,strength4}, it can be a 
good candidate for making novel strain devices~\cite{Neto}. One of the 
most remarkable properties induced by the strain in graphene is the 
appearance of the pseudo-magnetic fields and the corresponding pseudo-Landau 
level (pLLs)~\cite{pLL1,pLL2,pLL3,pLL4,pLL5,pLL6,pLL7,pLL8,pLL9}, 
which have been observed in several excellent 
experiments~\cite{exp1,exp2,exp3,exp4,exp5,exp6}. 

\begin{figure}[ht!]
\centering
\includegraphics[width=\linewidth]{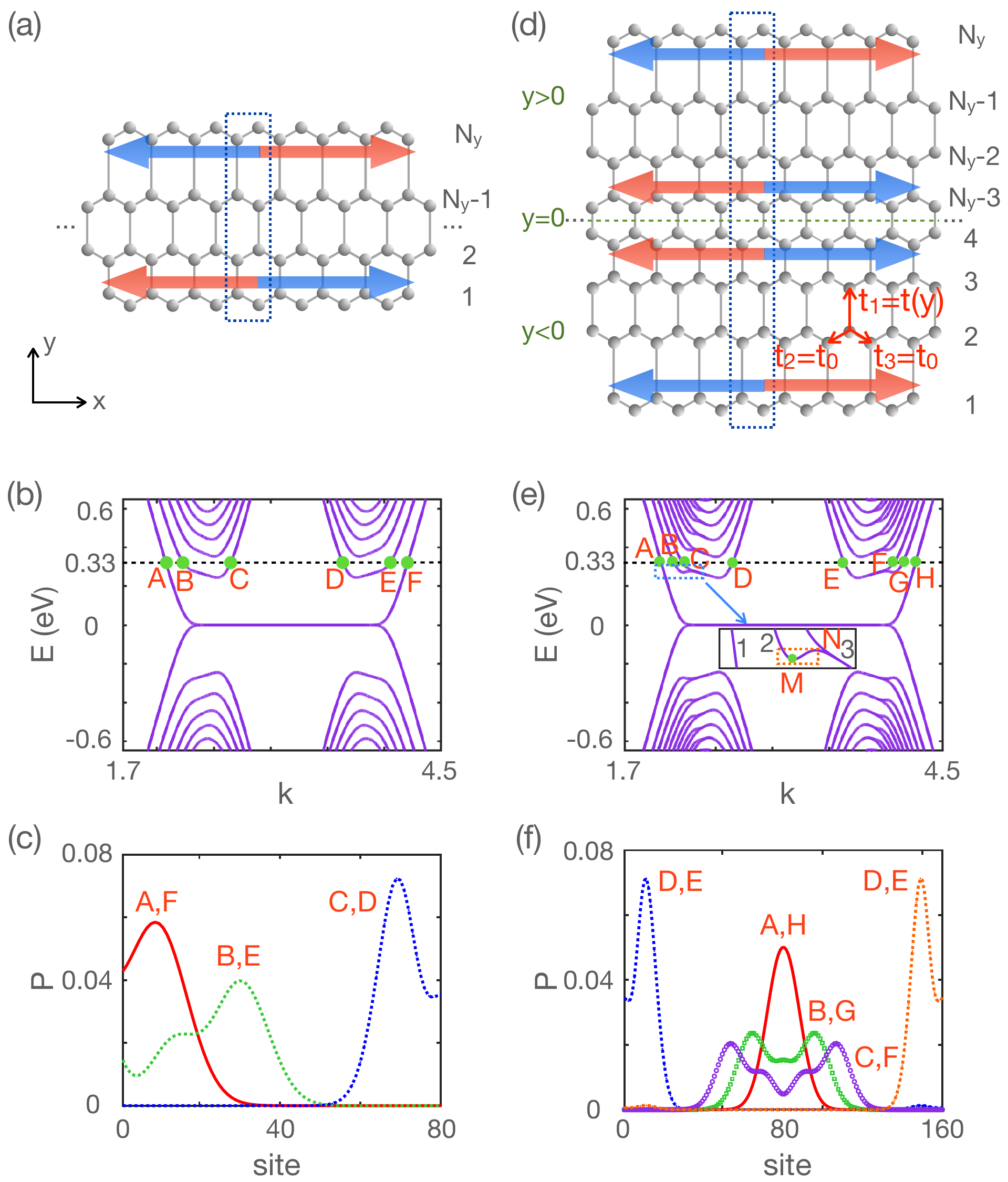}
\caption{(a) and (d) are the schematic diagrams of the ZGNR 
with MIS and SS, respectively. The blue dotted rectangle represents 
a primitive cell for the ZGNR, and the red (blue) arrows depict the 
current for $K$ ($K'$) valley. (b) and (c) ((e) and (f)) are 
energy bands and the distributions of wave functions for the 
ZGNR with MIS (SS).}
\label{fig:1}
\end{figure}

Previous work usually considers the model shown in 
Fig.~\ref{fig:1}(a)~\cite{pLL5,exp3,Franz}, where the ZGNR is stretched 
along the $y$ direction, and assuming the hopping coefficients 
are constant along the $x$ direction but decreasing successively 
from the lower edge to the upper one. Obviously, stretching 
increases the bond lengths, thus it will decrease the hopping 
coefficients which depend on the overlap integral of wave functions 
belonging to neighbouring sites. As a result, we call this strain 
pattern the monotone increasing strain (MIS), and if the hopping 
coefficients decrease linearly, a uniform perpendicular pseudo-magnetic field 
emerges and acts on the electrons locally, leading to the specific 
dispersive pLLs~\cite{pLL5,Franz}. Until now, the MIS 
configuration in GNRs has been well-studied and can be realized 
by various methods~\cite{pLL7,Sela1}. Even though the ZGNRs with MIS 
are charming, we suppose that another strain 
pattern shown in Fig.~\ref{fig:1}(d) may be easier to realize. 
It is clearly that the ZGNR with symmetrical structures in 
Fig.~\ref{fig:1}(d) can be viewed as two copies of the ZGNR in 
Fig.~\ref{fig:1}(a) where the lower copy ($y<0$) is the horizontal inversion
partner of the upper copy ($y>0$). The corresponding strain pattern we named 
the symmetrical strain (SS). It should be pointed out that even 
though we call it SS and the structure in Fig.~\ref{fig:1}(d) respects 
the inversion symmetry, the accurate symmetry between the lower and upper 
planes is not necessary. To be precise, what we need are the reverse  
strains between the lower and half planes, and without loss of generality, 
we use the SS to simplify our discussions.
We can expect that the ZGNR with SS may be more natural to 
fabricate in the suspended devices~\cite{SS1,SS2,SS3}. 
Simply speaking, if the ZGNR is stretched along the $y$ direction, 
and the $n_{y}=1$ and $n_{y}=N_{y}$ edges are fixed by probes or 
substrates, the SS pattern maybe form naturally. In Ref.~\cite{SS1}, 
a single graphene layer is placed over an array of pits etched in 
SiO$_{2}$/Si substrate. Continuously deformation can be tuned in 
this suspended device, producing regions with opposite signs of the pseudo-magnetic 
fields. In fact, what we need is the device naturally allowing 
for pseudo-magnetic fields changing sign, which occurs at the 
center of the ZGNR with reverse strains.

Polarized valley currents are predicted in the ZGNR with MIS~\cite{Franz}, 
so the meaningful question is what form should the valley currents present 
in the ZGNR with SS. Interestingly we found that except for the 
states propagating along the edges, emergent snake states can flow 
in the middle of the sample, which constitutes a new type of quantum 
valley Hall effect (QVHE). Even though this kind of snake 
states may be viewed as the bound states near the domain wall — 
1D ``line defect" along the $y=0$ line — between the lower and 
upper half parts, they are still fascinating because: (i) this 1D ``line 
defect" is natural and ``invisible", and we don't need to dope any 
real impurities; (ii) the reverse strain-induced snake states 
come from a fascinating phenomenon — the pseudo-magnetic fields 
have opposite directions between the lower and upper half parts, 
which is the result of the SS.

The paper is organized as follows. In Sec.~\ref{sec:edge}, 
we show the formation mechanism of the reverse strain-induced 
snake states. In Sec.~\ref{sec:valley}, we propose that the pure 
valley currents exist in the ZGNR with SS, which can be measured 
through the charge transport. In Sec.~\ref{sec:conductance}, we 
verify the existence of the snake states by the 
conductance of the two-terminal device calculated by the 
Landauer-B\"{u}ttiker formalism and nonequilibrium Green's function 
method, and discuss the robustness of the snake 
states. Sec.~\ref{sec:conclusions} is the conclusion. 

\section{The appearance of the reverse strain-induced snake states}\label{sec:edge}

The Hamiltonian in the tight-binding representation of the 
strained ZGNRs is $\mathcal{H}=\sum_{i}\varepsilon_{i}a_{i}^{\dagger}a_{i}-\sum_{<ij>}ta_{i}^{\dagger}a_{j}$, 
where $\varepsilon_{i}$ is the onsite energy,  
$a_{i}^{\dagger}$ and $a_{i}$ represent the creation and 
annihilation operators. For simplicity, we only consider 
the nearest-neighbor hopping, neglecting the next-nearest-neighbor 
hopping~\cite{Franz}. The strain pattern is contained in 
the hopping coefficient $t$. Since the strain is along the $y$ 
direction, we assume that the hopping coefficients along the 
$x$ direction, such as $t_{2}$ and $t_{3}$ in Fig.~\ref{fig:1}(d) 
are constant and set as the well-known hopping coefficient 
for the normal graphene $t_{0}=-2.75$~eV~\cite{Sun}. 
In the $y$ direction, the hopping coefficient $t(y)$ is dependent 
on the coordinate $y$ and we suppose that it is a linear function 
of $y$ in both upper and lower half planes. Because we assume 
that the positions of both edges ($n_{y}=1$ and $n_{y}=N_{y}$) are 
fixed, which can be realized in the suspended graphene, the SS 
configuration could be formed. In this situation, the strength of 
deformation is minimum (maximum) in the middle (edge) position. 
For simplicity, we suppose $t(y)$ is $t_{0}$ and $t_{0}(1-\eta)$ in 
the middle and edge position, respectively, where $\eta$ is an 
adjustable variable which reflects the strain strength. In the following 
calculations, we take $\eta=0.5$, which can be achieved in 
the experiments. First, we know that the graphene can sustain 
nondestructive reversible deformations up to high values 
$25\%$ to $27\%$~\cite{strength1,strength2,strength3,strength4}. 
Second, previous work has presented this relation 
$t_{y}=t_{0}\exp\left[-\beta\left(\ell/a_{0}-1\right)\right]$~\cite{strength4}, 
where $\ell$ is the carbon-carbon distance, $a_{0}=0.142$ nm 
is the equilibrium distance between neighboring carbon atoms, 
and $\beta\approx 3.37$ is the decay rate. As a result, if we 
take $\ell=1.25a_{0}$ for the deformation value $25\%$, we will get 
$t_{y}\approx 0.43t_{0}$, which suggests that the strain strength 
$\eta=0.5$ falls in the nondestructive reversible region. 

Fig.~\ref{fig:1}(b) shows the band structure of the ZGNR with 
MIS in Fig.~\ref{fig:1}(a) with $N_{y}=80$ and $\eta=0.5$. In our 
calculations, the hopping coefficients are supposed to be constant 
along the $x$ direction and to linearly decrease 
along the $y$ direction~\cite{pLL6}. As a result, the difference 
of $t_{y}$ between neighboring carbon atoms along the $y$ 
direction is approximately $0.6\%$. It is clearly that the 
dispersive pseudo-Landau levels form in the low energy regime, 
which is consistent with Ref.~\cite{pLL6, Franz}. In order to 
explore the probability distributions of the wave functions 
belonging to each band, we analysis the situations for selected 
$k$ points. Typical results are demonstrated in Fig~\ref{fig:1}(c), 
where we have chosen the $k$ points A-F at the intersections 
of the energy line $E=0.33$ eV and the lowest energy bands 
in Fig.~\ref{fig:1}(b). It is clearly shown that the wave functions 
associated with points A and C distribute near the lower and 
upper edges, respectively. As a comparison, the wave function 
of point B spreads into the bulk. We have checked that 
the wave functions of adjacent $k$ points near A, B, 
C have the similar localized properties with them. 
Additionally, the band structures and wave functions 
associated with D, E, F are the same with that of C, B 
and A, which is shown in Figs.~\ref{fig:1}(b) and (c). 

Fig.~\ref{fig:1}(e) shows the band structure of the ZGNR with 
SS in Fig.~\ref{fig:1}(d) with $N_{y}=160$ and $\eta=0.5$. 
The pseudo-magnetic field induced by the 
strain is about $49$T, which is estimated in Appendix~\ref{sec:A1}.
The most difference between the ZGNR with SS and the one with MIS 
is the former possesses reverse strains between 
the lower and half planes, which provides the possibility for the 
snake states discussed below. Similar to 
Fig.~\ref{fig:1}(b), the $k$ points A-H at the intersections of the 
energy line $E=0.33$ eV and the lowest energy bands are chosen 
in Fig.~\ref{fig:1}(e). Combined with the results in Fig.~\ref{fig:1}(f), 
we can see that the wave functions associated with points A and H 
(D and E) moves to the middle position (new edges). This result 
can be easily understood since the wave functions associated with 
point A (C) distribute near the lower (upper) edge which holds no 
strain strength (the max strain strength). In the ZGNR with SS 
in Fig.~\ref{fig:1}(d), the distributions follow the same rule, 
i.e., the wave functions associated with points A and H 
(D and E) distribute around the positions with no strain strength 
(the max strain strength). Similarly, the wave function associated 
with points B, G, C and F spreads into the bulk. In addition, 
from the partial enlarged diagram in Fig.~\ref{fig:1}(e) we can 
see that, the first pLL splits into band 2 and band 3 at point N 
due to the boundary condition~\cite{splitting}. Accordingly, a 
minimal point M emerge, and the surround band in the orange 
dotted rectangle will contribute additional edge states when 
the Fermi level across. 

\begin{figure}[ht!]
\centering
\includegraphics[width=\linewidth]{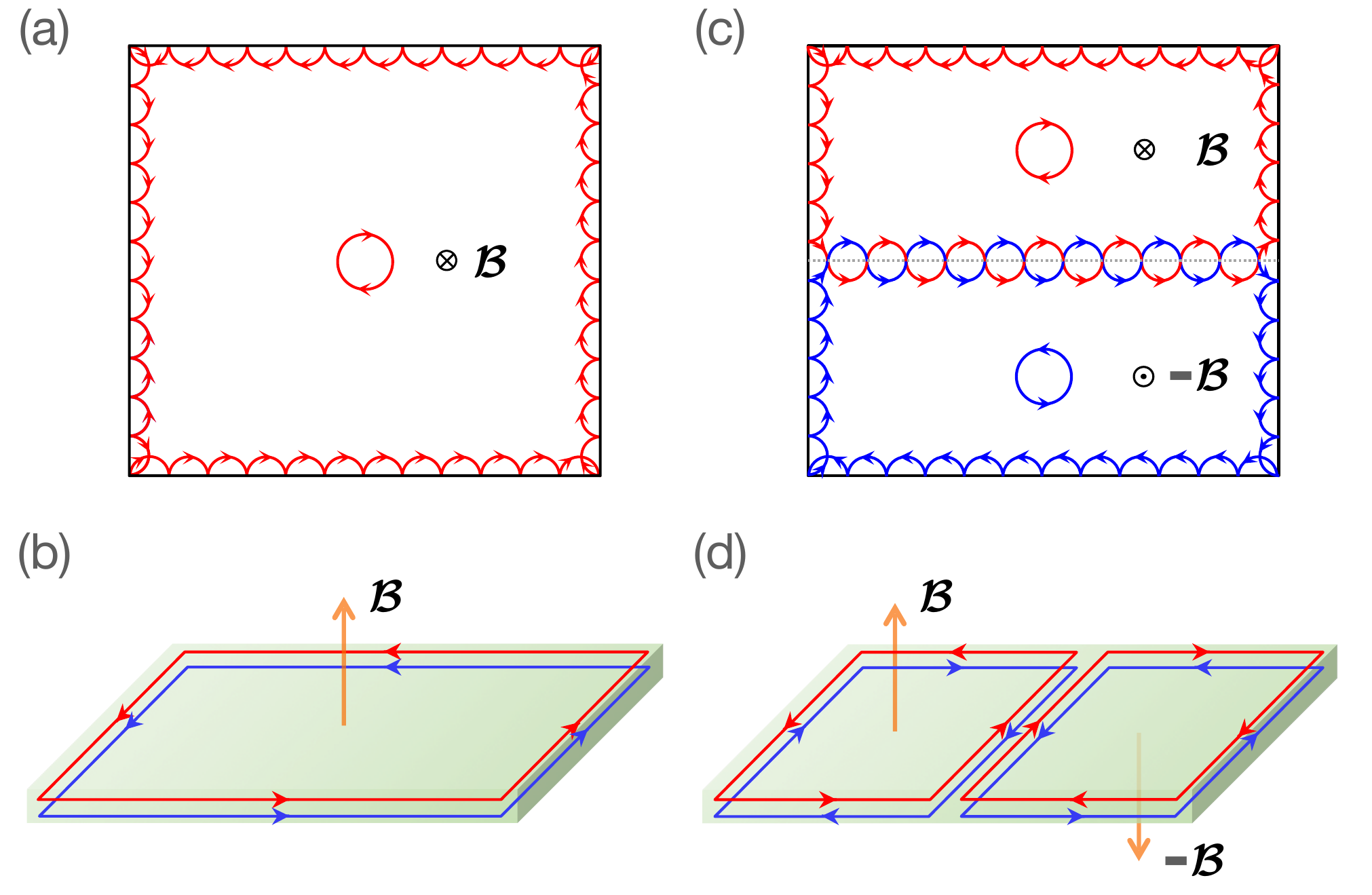}
\caption{(a) and (b) show the QH edge states, where 
the electrons take circular motions under the perpendicular 
magnetic field. (c) and (d) show the QVH edge states, where 
the electrons take circular motions under the perpendicular 
pseudo-magnetic fields. Because the electrons experience 
opposite Lorentz forces as they cross the zero field border line, 
the snake states propagate along the interface.}
\label{fig:2}
\end{figure}
%(a) and (b) show the QH edge states; (c) and (d) 
%show the QVH edge states which contain the middle ``edge 
%states".

It should be pointed out that the two valleys $K$ and $K'$ are 
decoupled in the low-energy limit making the study on the 
valley transport meaningful, as long as no scattering terms 
can connect the two valleys. As a result, if the Fermi energy 
lies between adjacent pLLs, the edge states with valley degree 
of freedom will dominate the transport. We compare the difference 
between edge states in our model and that in the normal quantum 
Hall (QH) states in Fig.~\ref{fig:2}. The key point is that 
the directions of the pseudo-magnetic fields, as well as the 
reverse strains, are opposite between the lower ($y<0$) and upper 
($y>0$) half planes. As a result, the electrons experience  
opposite Lorentz forces as they cross the zero field border line, 
resulting in the emergence of the snake states~\cite{Sela2,Egger,snake1,snake2} 
in the middle of the sample. In our setup, the snake states and 
the edge states co-participate the circular flow in both half planes, 
leading to the new type of QVH effect, which can be viewed as 
two copies of conventional QVH states.   

\section{Pure valley currents}\label{sec:valley}

\begin{figure}[ht!]
\centering
\includegraphics[width=\linewidth]{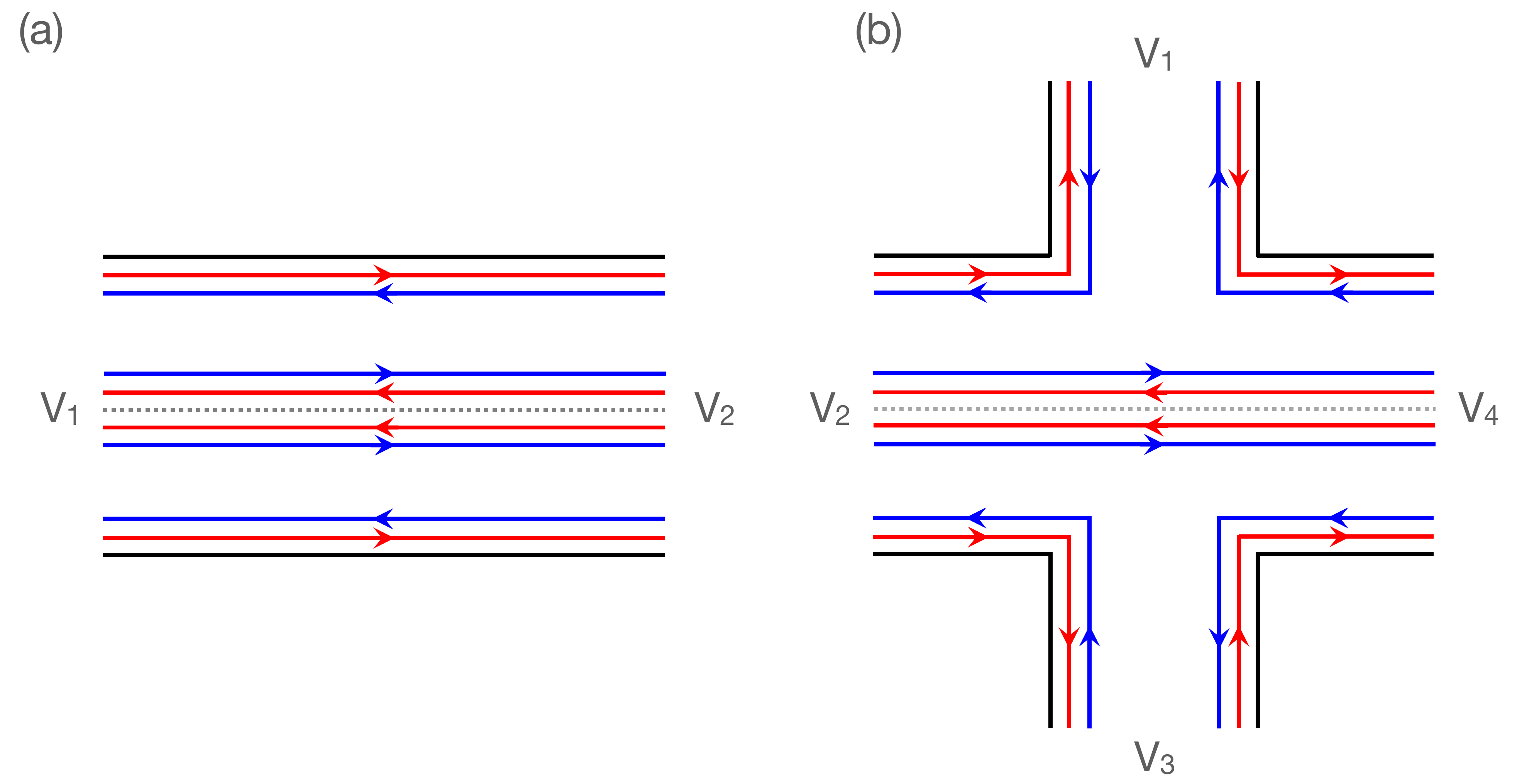}
\caption{Schematic diagram showing (a) two-terminal 
and (b) four-terminal measurement geometries. In 
(a) a charge current $I^{c}=4e^2V/h$ flows into 
the right lead. In(b) a valley current $I^{v}=e^2V/h$ 
flows into the right lead.}
\label{fig:3}
\end{figure}

The generation of pure valley currents has attracted lots of 
attention in the past few years. Several theoretical schemes, 
e.g., optical excitations~\cite{generation1}, quantum pumping~\cite{generation2,generation3}, 
cyclic strain deformations~\cite{generation4}, and applying 
AC bias~\cite{generation5}, have been proposed for realizing 
pure valley currents. In the laboratory, pure valley currents 
also have been observed in graphene superlattice~\cite{generation6} 
and graphene bilayers~\cite{generation7,generation8,generation9}. 
However, pure valley currents are very difficult to be observed 
in the monolayer graphene~\cite{Franz,Benjamin}.

Using similar methods in Ref.~\cite{Kane} in which pure 
spin currents are discussed, we use the Landauer-B\"{u}ttiker 
formula to discuss the valley currents in two and four terminal 
devices shown in Fig.~\ref{fig:3}. Different from the usual 
QH and quantum spin Hall effect (QSHE) where the 1D 
conducting states localizing in the edges, the snake 
states exist in our device and contribute to transport. According 
to the B\"{u}ttiker formula~\cite{Buttiker}, the current in $i$-th 
terminal in the equilibrium is 

\begin{align}
I^{\sigma}_{i}=\frac{e^2}{h}\sum_{j\neq i}\left(T^{\sigma}_{ji}V_{i}-T^{\sigma}_{ij}V_{j}\right),   
\end{align}
where $V_{i}$ is the voltage in the $i$-th terminal and $T^{\sigma}_{ij}$ 
is the transmission coefficient for valley $\sigma$ 
($\sigma=K,K'$) current between the $i$-th 
and $j$-th terminals. Naturally, the charge current and valley 
current can be defined by $I^{c}_{i}=I^{K}_{i}+I^{K'}_{i}$ 
and $I^{v}_{i}=I^{K}_{i}-I^{K'}_{i}$.

For the two-terminal device shown in Fig.~\ref{fig:3}(a), 
$T^{\sigma}_{12}=T^{\sigma}_{21}=2$, thus we get 
$I^{K}_{2}=\left(2V_{1}-2V_{2}\right)e^2/h$ and 
$I^{K'}_{2}=\left(2V_{1}-2V_{2}\right)e^2/h$ 
for terminal 2. If we take $V_{1}=V/2$ and $V_{2}=-V/2$, 
then $I^{K}_{2}=2e^2V/h$, $I^{K'}_{2}=2e^2V/h$. 
As a result, $I^{c}_{2}=4e^2V/h$, $I^{v}_{2}=0$. 
This suggests that the pure valley currents cannot exist 
in the two-terminal device. Next we study the four-terminal 
device shown in Fig.~\ref{fig:3}(b) to see 
whether the pure valley currents can survive. In this case, 
the transmission coefficients for $K$ valley currents 
are $T^{K}_{14}=T^{K}_{21}=T^{K}_{23}=T^{K}_{34}=1$, 
$T^{K}_{42}=2$ and $0$ otherwise; the transmission 
coefficients for $K'$ valley currents are 
$T^{K'}_{12}=T^{K'}_{32}=T^{K'}_{41}=T^{K'}_{43}=1$, 
$T^{K'}_{24}=2$ and $0$ otherwise. Therefore, 
we can get $I^{c}_{4}=\left(V_{1}+2V_{2}+V_{3}-4V_{4}\right)e^2/h$ 
and $I^{v}_{4}=\left(-V_{1}+2V_{2}-V_{3}\right)e^2/h$. 
If we take $V_{1}=-V/2$, $V_{2}=V/4$, $V_{3}=V_{4}=0$,  
we have $I^{c}_{4}=0$ but $I^{v}_{4}=e^2V/h$, which suggest 
that we can obtain the pure valley currents in terminal 4. 
In other words, the pure valley currents can be measured 
through the charge transport in a mesoscopic system 
as shown in Fig.~\ref{fig:3}. It should be pointed out that 
this is one simple and elegant way to realize the pure valley 
currents in the monolayer graphene systems.

\section{Conductances from the edge states and snake states}\label{sec:conductance}

\begin{figure}[ht!]
	\centering
	\includegraphics[width=\linewidth]{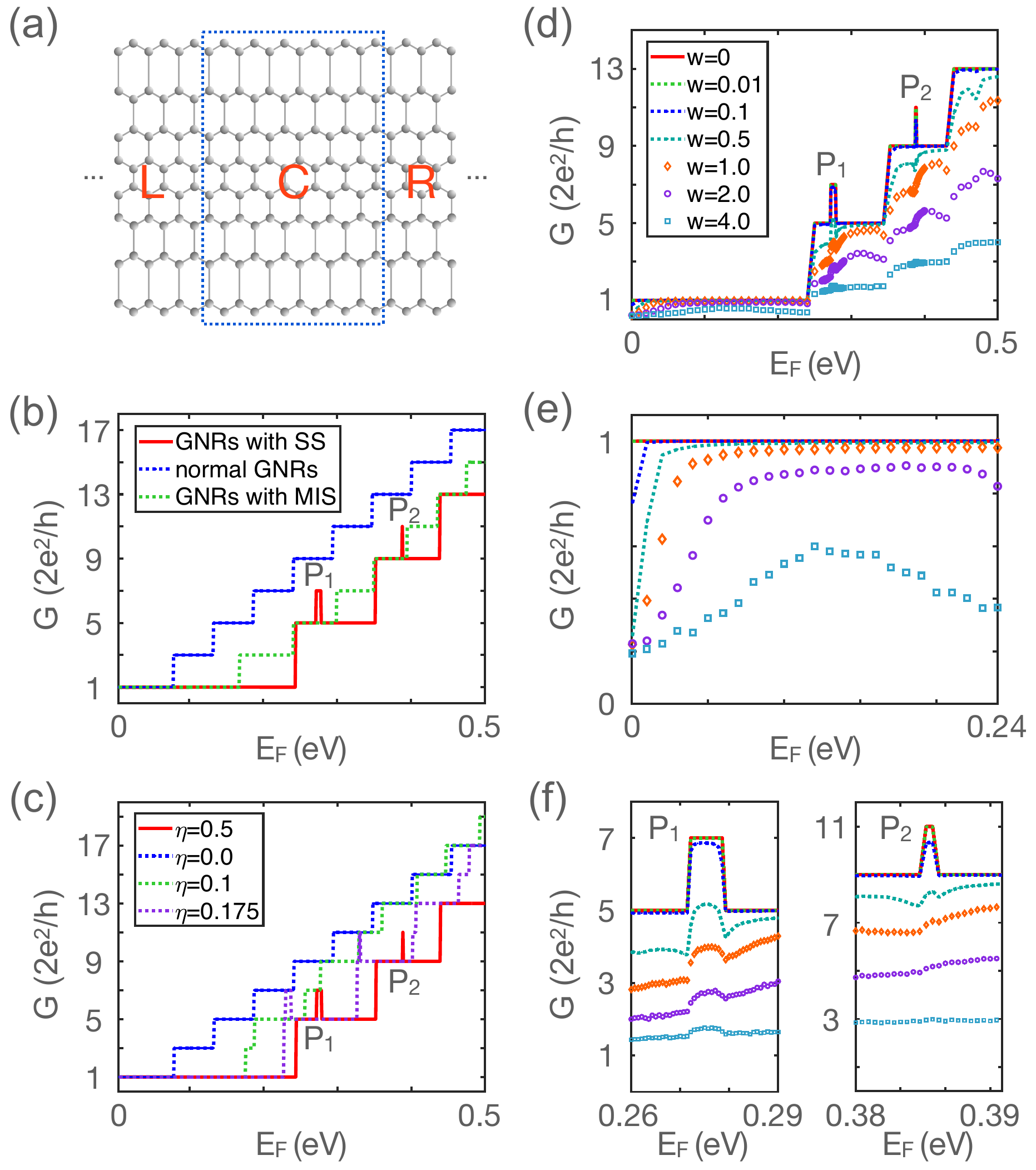}
	\caption{Conductances versus Fermi energy for the strained ZGNRs. 
	(a) is the schematic diagram of the two-terminal device. (b) 
	shows the conductance plateaus for the ZGNR without strain, 
	ZGNR with SS and ZGNR with MIS, respectively. (c) illustrates 
	the conductances of the ZGNR with SS for different strain strength. 
	(d)-(f) show the influences of Anderson disorders 
	on conductances with (e) and (f) from (d).}
	\label{fig:4}
\end{figure}

Next we use the nonequilibrium Green's function method to 
calculate the conductance of the strained ZGNR shown in 
Fig.~\ref{fig:4}(a), where $L$ and $R$ denote the left and 
right semi-inﬁnite long leads, and $C$ represents the central part with 
the width $N_{y}=160$ and length $N_{x}=60$ (60 primitive cells 
in the $x$ direction). It should be pointed 
out that the central part and the leads are all stretched along the 
transverse direction, thus the snake states could 
survive in the whole system. Since our calculations based on the tight binding 
model, the central part and two leads are all described by the Hamiltonian 
$\mathcal{H}=\sum_{i}\varepsilon_{i}a^{\dagger}_{i}a_{i}-\sum_{<ij>}ta^{\dagger}_{i}a_{j}$, 
even though each part may possess different onsite energies $\varepsilon_{i}$.
The linear conductance of the strained ZGNR is $G=\lim_{V\to0}dI/dV$, 
where the current $I$ is obtained by the Landauer-B\"{u}ttiker formula~\cite{Ren,Sun}: 
$I=\left(2e/h\right)\int d\varepsilon T_{LR}\left(\varepsilon\right)\left[f_{L}\left(\varepsilon\right)-f_{R}\left(\varepsilon\right)\right]$. 
Here $f_{\alpha}=1/\left\{\exp\left[\left(\varepsilon-eV_{\alpha}\right)/k_{B}T\right]+1\right\}$ 
is the Fermi function of the $\alpha$ $\left(\alpha=L,R\right)$ lead, 
and $T_{LR}\left(\varepsilon\right)=\textrm{Tr}\left(\boldsymbol{\Gamma}_{L}\boldsymbol{G}^{r}\boldsymbol{\Gamma}_{R}\boldsymbol{G}^{a}\right)$ is the transmission coefficient with the linewidth functions 
$\boldsymbol{\Gamma}_{\alpha}=i\left[\boldsymbol{\Sigma}_{\alpha}^{r}\left(\varepsilon\right)-\boldsymbol{\Sigma}_{\alpha}^{a}\left(\varepsilon\right)\right]$ and the self-energy 
$\boldsymbol{\Sigma}^{r/a}_{\alpha}$. The retarded and advanced Green's function is 
$\boldsymbol{G}^{r}\left(\varepsilon\right)=1/\left[\varepsilon-\boldsymbol{H}_{cen}-\boldsymbol{\Sigma}_{L}^{r}-\boldsymbol{\Sigma}_{R}^{r}\right]$ and $\boldsymbol{G}^{a}\left(\varepsilon\right)=\left[\boldsymbol{G}^{r}\left(\varepsilon\right)\right]^{\dagger}$, 
respectively, where $\boldsymbol{H}_{cen}$ is the Hamiltonian of the 
central region. In the following numerical calculations, we take the 
hopping energy $t_0\approx2.75$ eV, and since the $t_0$ corresponds 
to $10^{4}$ K, we can safely set the temperature to zero in 
our calculations~\cite{Sun}. 

We first study the clean strained ZGNRs. Fig.~\ref{fig:4}(b) compares 
three stain patterns: the blue dash-dot line shows the conductance 
without strain, which is quantized and exhibits a series of equidistant 
plateaus due to the transverse sub-bands of the ZGNR with finite width. 
The green dot line corresponds to the conductance of the ZGNR with 
MIS. It is clear that due to the formation of the pLLs, the width 
of the lower plateaus increases. Furthermore, the step interval of the plateaus 
maintains $4e^2/h$ which is attributed to the states located in the two 
edges. After we introduce the SS, the step interval of the plateaus increases 
to $8e^2/h$, which implies the formation of the snake states 
in the middle of the sample. It should be pointed out that two peaks 
$P_{1}$ and $P_{2}$ emerge on the second and third plateaus due to 
the energy bands inside the orange dot rectangle in Fig.~\ref{fig:1}(e). 
More discussions on $P_{1}$ and $P_{2}$ can be found in 
Appendix~\ref{sec:A2}. In fact, $P_{1}$ and $P_{2}$ 
are two narrow plateaus which is clearly shown in Fig.~\ref{fig:4}(f). 
From Fig.~\ref{fig:1}(e) we can see that the points similar with the 
emergent minimum point N are more difficult to appear on higher pLLs, 
thus the strengths of the similar narrow plateaus become smaller with 
increasing $E_{F}$.

In fact, the strain strength plays the role of the pseudo-magnetic 
fields, contributing to the pLLs and the quantized valley Hall conductance. 
Therefore, we plot Fig.~\ref{fig:4}(c) to illustrate the conductance with 
the change in strain, which shows the evolution of step interval from 
$4e^2/h$ to $8e^2/h$. Blue dot line corresponds to the ZGNR without strain, 
and according to the above analysis, the step interval is $4e^2/h$. The green 
dot line, purple dot line and the red solid line present the conductances 
versus the increase of strain. It is clearly shown that along with the 
increase of strain, the plateaus at $6e^2/h$, $14e^2/h$, $22e^2/h$, $30e^2/h$ 
become narrower. Until they disappear, the step interval between 
adjacent plateaus changes from $4e^2/h$  to $8e^2/h$. Meanwhile, $P_{1}$ 
and $P_{2}$ appear gradually because of the formations of new edge states. 

Next we examine the effect of disorder on the snake states. 
In the following calculations, we suppose that the disorder only exists 
in the central region, and in the presence of disorder, we take $500$ random 
configurations and calculate the average values of the conductances. Since 
this procedure has high computation cost, we cannot take the same number of 
$E_{F}$ as that in Fig.~\ref{fig:4}(b) and (c). However, to avoid missing 
information related to $P_{1}$ and $P_{2}$, we take more number of $E_{F}$ 
around them in the calculations. The detailed results are illustrated 
in Fig.~\ref{fig:4}(d) and the drawing of partial enlargement 
of the first plateau, $P_{1}$ and $P_{2}$ are shown in 
Fig.~\ref{fig:4}(e) and (f), respectively. The results can 
be summarized as follows: (i) when the strength of disorder is very weak, e.g., 
$w\leq 0.1$ eV, all plateaus, except for $P_{1}$ and $P_{2}$, maintain 
well. $P_{1}$ and $P_{2}$ are not robust because the range 
of band structure inside the orange dot rectangle in Fig.~\ref{fig:1}(e) 
is small. Moreover, this type of peaks will become more fragile on 
higher plateaus. (ii) When the strength of disorder 
increase, e.g., $w=0.5$ eV, only the first plateau is robust.
It is noted that there is a minimum of conductance in the vicinity of $E_{F}=0$ 
in Fig.~\ref{fig:4}(d) and (e), which may be explained 
by the Anderson localization~\cite{Mucciolo}. Since the number 
of propagating channels in the narrow ribbons is very small 
near $E_{F}=0$, the GNR can be viewed as a quasi-1D system. 
As a result, the conductance dip will appear due to the Anderson 
localization if the length of the GNR is longer than the localization 
length. The physical reason why the conductance plateaus is 
not so robustness is the snake states are easy to become 
hybrid due to the close distance between them in the real space. 
Therefore, we need to explore schemes preventing the hybridization 
and obtain more robust snake states in the middle of 
the sample.

\section{Conclusions}\label{sec:conclusions}

The valley degree of freedom attributed to the non-equivalent Dirac 
points $K$ and $K'$ is well-known for decades. Furthermore, strain 
can be used to tailor the band structures and properties of the ZGNRs, 
and the most charming phenomenon is the emergent pseudo-magnetic fields 
and the corresponding pLLs. In order to explore new types of pure 
valley currents, we design the strained ZGNR which can support the 
snake states propagating in the middle of the sample. We point out that 
the reverse strains between the lower and half planes is the essential 
ingredient for the formation of the snake states. Based on our 
design, we can use a four-terminal device to realize the long-thought 
pure valley currents in monolayer graphene systems. Furthermore, we 
calculate the conductance of the two-terminal device, obtaining 
the $8e^2/h$ step interval between neighbouring plateaus, which 
indicates the formation of the snake states in the middle 
of the sample. By introducing the Anderson impurities, we find that 
the first conductance plateau is more robust than higher plateaus. 
One obstacle to the robustness of the snake states 
is they are easy to become hybrid due to the close distance between 
them in the real space. One of the next aim is to explore schemes 
preventing the hybridization and obtain more robust snake 
states in the middle of the sample.

\begin{acknowledgments}

The authors would like to thank Qing-feng Sun, Shuai Zhao, 
Chui-zhen Chen and Shu-feng Zhang for helpful discussions. We 
are grateful to the National Natural Science Foundation of China 
(Nos. 11774093, 11104075), Science and Technology Commission of 
Shanghai Municipality (No. 16ZR1409800), Natural Science Foundation 
of Shanghai (Grant No. 21JC1402300) and Director's Fund of Key 
Laboratory of Polar Materials and Devices, Ministry of Education.

\end{acknowledgments}

\begin{appendix}

\section{The value of the pseudo-magnetic field}\label{sec:A1}

\begin{figure}[htb]
\centering
\includegraphics[width=\linewidth]{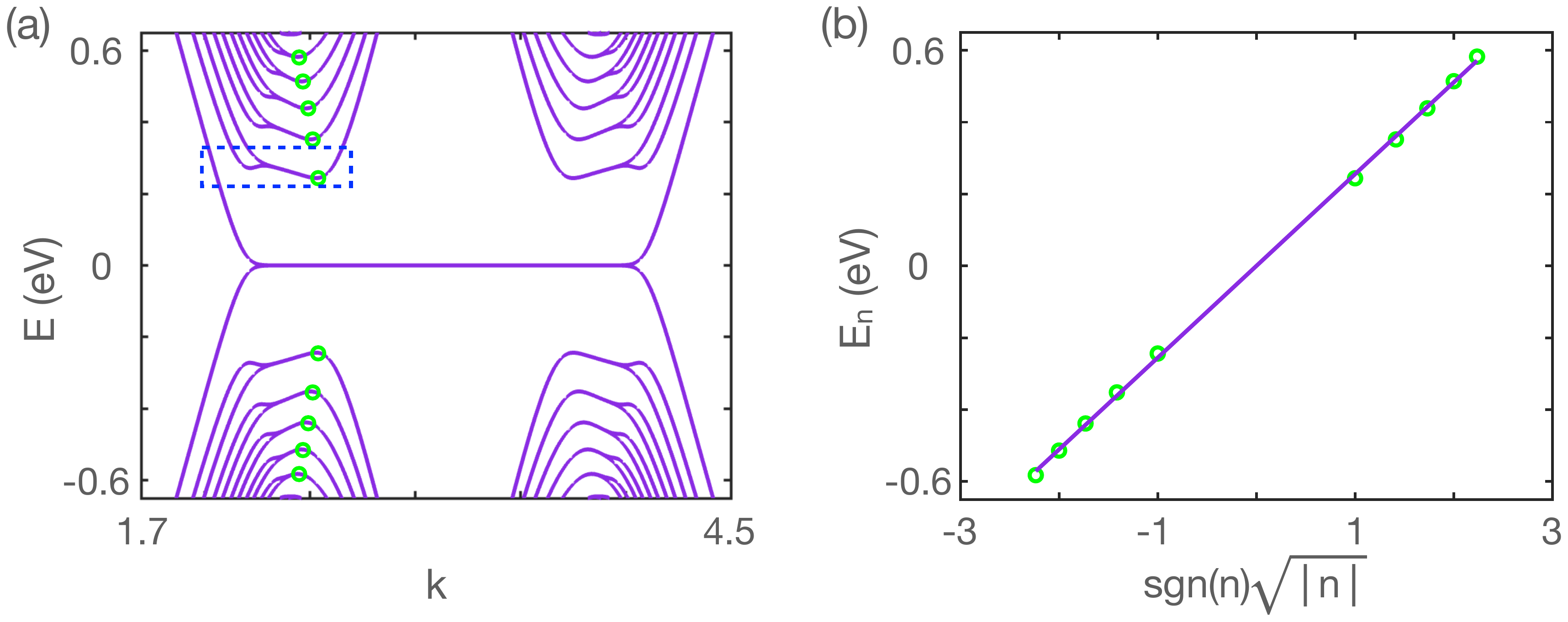}
\caption{(a) The band structure adopted from 
Fig.~\ref{fig:1}(e). (b) The comparison between 
the energy points marked by green circles in (a) and the Landau 
levels generated by the real magnetic field.}
\label{fig:A1}
\end{figure}

The value of the pseudo-magnetic field in our model can 
be estimated by the method used in Ref.~\cite{pLL6}. 
Fig.~\ref{fig:A1}(a) is the band structure adopted from 
Fig.~\ref{fig:1}(e). We select several energy points marked 
by green circles and compare them with the Landau levels generated 
by the real magnetic fields $E_{n}=v_{F}\sqrt{2e\hbar B}\mathrm{sgn}(n)\sqrt{|n|}$, 
where $n$ is the index of the Landau levels and $v_{F}\approx 10^{6}\mathrm{m/s}$ 
is the Fermi velocity. We can read from the slope in Fig.~\ref{fig:A1}(b) that 
the pseudo-magnetic field $B\approx 49\mathrm{T}$. It should be 
pointed out according to the relations:

\begin{align}
A_{x}=&\frac{c}{2ev_{F}}\left(t_{2}+t_{3}-2t_{1}\right)=\frac{c}{ev_{F}}\left(t_{0}-t(y)\right), \nonumber \\
A_{y}=&\frac{\sqrt{3}c}{2ev_{F}}\left(t_{3}-t_{2}\right)=\frac{\sqrt{3}c}{2ev_{F}}\left(t_{0}-t_{0}\right)=0,
\end{align}
the magnetic field $\mathbf{B}=\nabla\times\mathbf{A}$ is a constant 
field along the $z$ direction since we have assumed that $t(y)$ is 
a linear function of $y$ in both upper and lower half planes.

\section{Discussions on $P_{1}$ and $P_{2}$}\label{sec:A2}

\begin{figure}[htb]
\centering
\includegraphics[width=\linewidth]{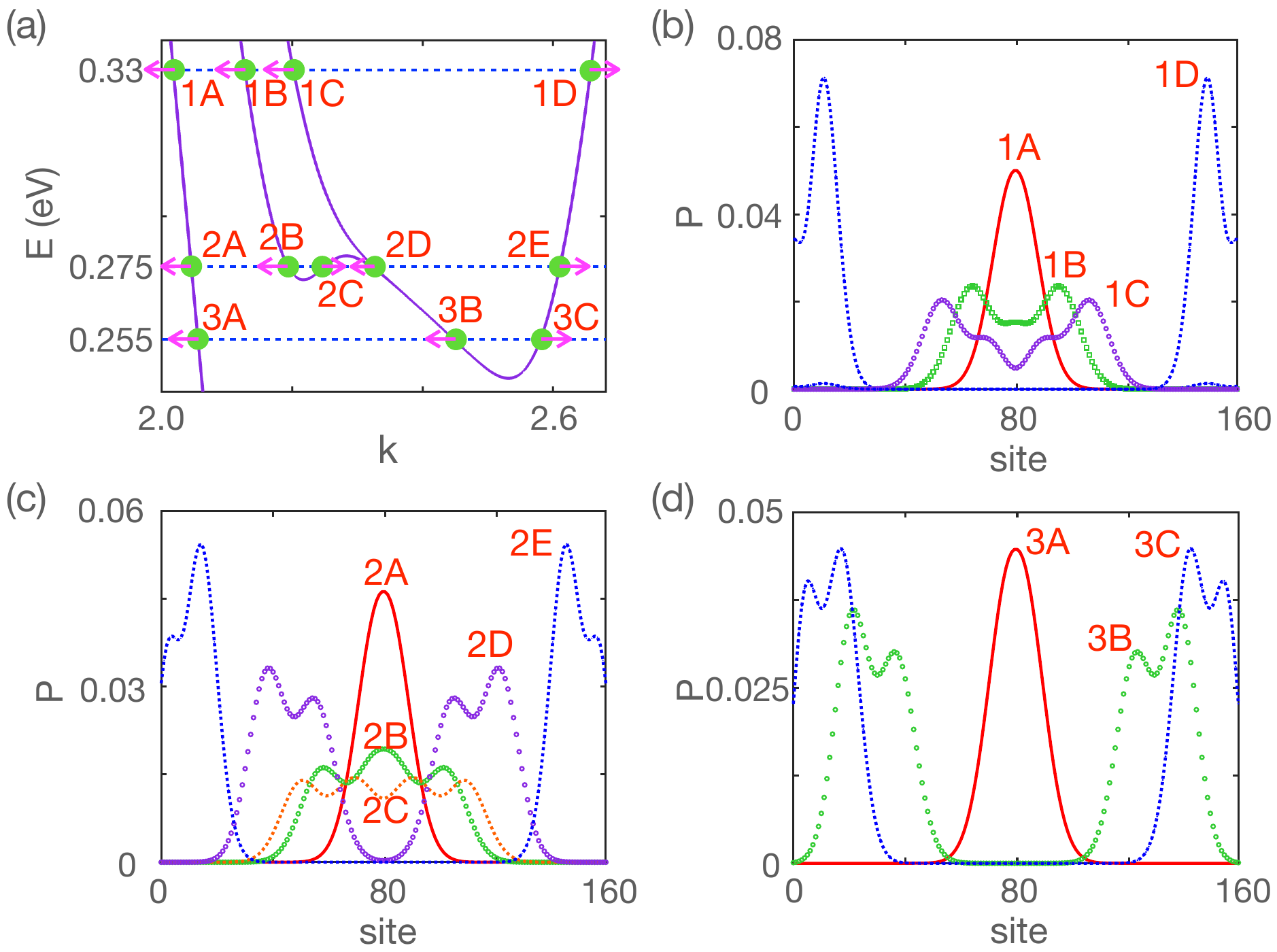}
\caption{Spatial distributions of the eigenstates belong 
to $E=0.33$ eV, $E=0.275$ eV and $E=0.255$ eV, respectively.}
\label{fig:A2}
\end{figure}

Fig.~\ref{fig:A2}(a) is the partial enlarged 
picture of the bands in the blue dashed box in Fig.~\ref{fig:A1}(a). In order 
to show the reason why the extra conductance $2e^2/h$ ($4e^2/h$ 
if spin is considered) emerges due 
to $P_{1}$ more clearly, we plot the spatial distributions of the 
eigenstates belong to $E=0.33$~eV, $E=0.275$~eV and $E=0.255$~eV 
in Figs.~\ref{fig:A2}(b)-(d), respectively. It is clear that $P_{1}$ emerges 
around $E=0.275$~eV because of the extra intersection points 2B 
and 2C, which contributes the conductance $2e^2/h$. From Fig.~\ref{fig:A2}(c) 
we can see that the wave functions of 2B and 2C are localized 
around the middle region of the sample. The emergence of $P_{2}$ 
can be analysed in the same way.

It should be pointed out that, one of the main difference between 
the pseudo-Landau levels and the conventional landau levels is the 
former are dispersive because of the interaction between the bulk 
states and edge states. As a result, crossing with the bulk states 
belonging to the pseudo-Landau levels can always appear when 
we continuously tune the Fermi energy, e.g., the intersection point 
3B in Fig.~\ref{fig:A2}. Furthermore, due to the interaction between the bulk 
states and edge states, the localization strength of certain states 
is not that strong. As we know, the robustnesses of the 
bulk states, edge states and snake states are different, 
as a result, the stability of the conductance plateaus needs to be 
studied further.

\end{appendix}

\bibliographystyle{apsrev4-1}

\end{document}